\documentclass[11pt,a4paper]{article}

\usepackage{jheppub}
\usepackage{amsmath}
\usepackage{axodraw}
\usepackage{bbm}

\allowdisplaybreaks[1]

\DeclareMathOperator{\Li}{Li}

\newcommand{\thinlinewidth}{.6}
\newcommand{\thicklinewidth}{2.0}

\newcommand{\pCtwoonefoura}[0]{
\parbox{100pt}{\makebox(100,40){\hspace{-15pt}
\begin{picture}(0,0)
\SetScale{1}
 \SetWidth{\thicklinewidth}
 \Line(-15,-15)(0,-15) \Line(-15,15)(0,15) \Line(-15,15)(-15,-15)
 \Line(0,-15)(15,-15)  \Line(0,15)(15,15)  \Line(15,15)(40,15)
 \Line(15,-15)(40,-15)
 \SetWidth{\thinlinewidth}
 \Line(-30,15)(-15,15) \Line(-30,-15)(-15,-15) \Oval(15,0)(15,15)(0)
 \ArrowLine(-30,15)(-15,15)   \Text(-38,15)[]{$p_1$}
 \ArrowLine(-30,-15)(-15,-15) \Text(-38,-15)[]{$p_2$}
 \ArrowLine(25,15)(40,15)   \Text(49,15)[]{$p_3$}
 \ArrowLine(25,-15)(40,-15) \Text(49,-15)[]{$p_4$}
\end{picture}
}}}

\newcommand{\pCtwoonefourb}[0]{
\parbox{100pt}{\makebox(100,40){\hspace{-15pt}
\begin{picture}(0,0)
 \SetWidth{\thicklinewidth}
 \Line(-15,-15)(0,-15) \Line(-15,15)(0,15) \Line(-15,15)(-15,-15)
 \Line(0,-15)(15,-15)  \Line(0,15)(15,15)  \Line(15,15)(40,15)
 \Line(15,-15)(40,-15)
 \SetWidth{\thinlinewidth}
 \Line(-30,15)(-15,15) \Line(-30,-15)(-15,-15) \Oval(15,0)(15,15)(0)
 \ArrowLine(-30,15)(-15,15)   \Text(-38,15)[]{$p_1$}
 \ArrowLine(-30,-15)(-15,-15) \Text(-38,-15)[]{$p_2$}
 \ArrowLine(25,15)(40,15)   \Text(49,15)[]{$p_3$}
 \ArrowLine(25,-15)(40,-15) \Text(49,-15)[]{$p_4$}
 \GCirc(-15,0){3}{0}
\end{picture}
}}}

\newcommand{\pDonesevenfoura}[0]{
\parbox{100pt}{\makebox(100,40){\hspace{5pt}
\begin{picture}(0,0)
\SetScale{1}
 \SetWidth{\thinlinewidth}
 \Line(-40,15)(-15,15) \Line(-15,15)(0,15) \Line(0,15)(15,15)
 \Line(15,-15)(30,-15) \Oval(-15,0)(15,15)(0)
 \SetWidth{\thicklinewidth}
 \Line(-40,-15)(-15,-15) \Line(15,15)(30,15) \Line(15,15)(15,-15)
 \Line(-15,-15)(0,-15) \Line(0,-15)(15,-15)
 \SetWidth{\thinlinewidth}
 \ArrowLine(-40,15)(-25,15)   \Text(-48,15)[]{$p_1$}
 \ArrowLine(-25,-15)(-40,-15) \Text(-48,-15)[]{$p_3$}
 \ArrowLine(15,15)(30,15)   \Text(39,15)[]{$p_4$}
 \ArrowLine(30,-15)(15,-15) \Text(39,-15)[]{$p_2$}
\end{picture}
}}}

\newcommand{\pDonesevenfourb}[0]{
\parbox{100pt}{\makebox(100,40){\hspace{5pt}
\begin{picture}(0,0)
\SetScale{1}
 \SetWidth{\thinlinewidth}
 \Line(-40,15)(-15,15) \Line(-15,15)(0,15) \Line(0,15)(15,15)
 \Line(15,-15)(30,-15) \Oval(-15,0)(15,15)(0)
 \SetWidth{\thicklinewidth}
 \Line(-40,-15)(-15,-15) \Line(15,15)(30,15) \Line(15,15)(15,-15)
 \Line(-15,-15)(0,-15) \Line(0,-15)(15,-15)
 \SetWidth{\thinlinewidth}
 \ArrowLine(-40,15)(-25,15)   \Text(-48,15)[]{$p_1$}
 \ArrowLine(-25,-15)(-40,-15) \Text(-48,-15)[]{$p_3$}
 \ArrowLine(15,15)(30,15)   \Text(39,15)[]{$p_4$}
 \ArrowLine(30,-15)(15,-15) \Text(39,-15)[]{$p_2$}
 \GCirc(15,0){3}{0}
\end{picture}
}}}

\newcommand{\pDoneeighttwoa}[0]{
\parbox{100pt}{\makebox(100,40){\hspace{-5pt}
\begin{picture}(0,0)
\SetScale{1}
 \SetWidth{\thinlinewidth}
 \Line(-35,15)(-15,15) \Line(-15,15)(15,15)
 \Line(15,-15)(35,-15) \Line(-15,-15)(15,15) \Line(-15,-15)(-15,15)
 \SetWidth{\thicklinewidth}
 \Line(-35,-15)(-15,-15) \Line(15,15)(35,15) \Line(15,15)(15,-15)
 \Line(-15,-15)(15,-15)
 \SetWidth{\thinlinewidth}
 \ArrowLine(-35,15)(-20,15)   \Text(-43,15)[]{$p_1$}
 \ArrowLine(-20,-15)(-35,-15) \Text(-43,-15)[]{$p_3$}
 \ArrowLine(20,15)(35,15)   \Text(44,15)[]{$p_4$}
 \ArrowLine(35,-15)(20,-15) \Text(44,-15)[]{$p_2$}
\end{picture}
}}}

\newcommand{\pDoneeighttwob}[0]{
\parbox{100pt}{\makebox(100,40){\hspace{-5pt}
\begin{picture}(0,0)
\SetScale{1}
 \SetWidth{\thinlinewidth}
 \Line(-35,15)(-15,15) \Line(-15,15)(15,15)
 \Line(15,-15)(35,-15) \Line(-15,-15)(15,15) \Line(-15,-15)(-15,15)
 \SetWidth{\thicklinewidth}
 \Line(-35,-15)(-15,-15) \Line(15,15)(35,15) \Line(15,15)(15,-15)
 \Line(-15,-15)(15,-15)
 \SetWidth{\thinlinewidth}
 \ArrowLine(-35,15)(-20,15)   \Text(-43,15)[]{$p_1$}
 \ArrowLine(-20,-15)(-35,-15) \Text(-43,-15)[]{$p_3$}
 \ArrowLine(20,15)(35,15)   \Text(44,15)[]{$p_4$}
 \ArrowLine(35,-15)(20,-15) \Text(44,-15)[]{$p_2$}
 \GCirc(0,-15){3}{0}
\end{picture}
}}}

\newcommand{\pEthreethreethreea}[0]{
\parbox{100pt}{\makebox(100,40){\hspace{-5pt}
\begin{picture}(0,0)
\SetScale{1}
 \SetWidth{\thinlinewidth}
 \Line(-35,15)(-15,15) \Line(-15,15)(0,15) \Line(0,15)(15,15)
 \Line(15,-15)(35,-15)  \Line(-15,-15)(-15,15)
 \Line(-15,-15)(15,-15) \Line(15,15)(15,-15)
 \SetWidth{\thicklinewidth}
 \Line(-35,-15)(-15,-15) \Line(15,15)(35,15) \Line(-15,-15)(15,15)
 \SetWidth{\thinlinewidth}
 \ArrowLine(-35,15)(-20,15)   \Text(-43,15)[]{$p_1$}
 \ArrowLine(-20,-15)(-35,-15) \Text(-43,-15)[]{$p_4$}
 \ArrowLine(20,15)(35,15)   \Text(44,15)[]{$p_3$}
 \ArrowLine(35,-15)(20,-15) \Text(44,-15)[]{$p_2$}
\end{picture}
}}}

\newcommand{\pEthreethreethreeb}[0]{
\parbox{100pt}{\makebox(100,40){\hspace{-5pt}
\begin{picture}(0,0)
\SetScale{1}
 \SetWidth{\thinlinewidth}
 \Line(-35,15)(-15,15) \Line(-15,15)(0,15) \Line(0,15)(15,15)
 \Line(15,-15)(35,-15)  \Line(-15,-15)(-15,15)
 \Line(-15,-15)(15,-15) \Line(15,15)(15,-15)
 \SetWidth{\thicklinewidth}
 \Line(-35,-15)(-15,-15) \Line(15,15)(35,15) \Line(-15,-15)(15,15)
 \SetWidth{\thinlinewidth}
 \ArrowLine(-35,15)(-20,15)   \Text(-43,15)[]{$p_1$}
 \ArrowLine(-20,-15)(-35,-15) \Text(-43,-15)[]{$p_4$}
 \ArrowLine(20,15)(35,15)   \Text(44,15)[]{$p_3$}
 \ArrowLine(35,-15)(20,-15) \Text(44,-15)[]{$p_2$}
 \GCirc(0,0){3}{0}
\end{picture}
}}}

\newcommand{\nAfoursixthreea}[0]{
\parbox{100pt}{\makebox(100,40){\hspace{-15pt}
\begin{picture}(0,0)
\SetScale{1}
 \SetWidth{\thicklinewidth}
 \Line(25,15)(25,-15)
 \Line(25,-15)(40,-15) \Line(25,15)(40,15)
 \SetWidth{\thinlinewidth}
 \Line(-15,-15)(0,-15) \Line(-15,15)(25,15)
 \Line(0,-15)(15,-15) \Line(-30,-15)(25,-15)
 \Line(-15,-15)(-2,-2) \Line(2,2)(15,15)
 \Line(-15,15)(15,-15)
 \ArrowLine(-30,15)(-15,15)   \Text(-38,15)[]{$p_1$}
 \ArrowLine(-30,-15)(-15,-15) \Text(-38,-15)[]{$p_2$}
 \ArrowLine(28,15)(40,15)   \Text(49,15)[]{$p_3$}
 \ArrowLine(28,-15)(40,-15) \Text(49,-15)[]{$p_4$}
\end{picture}
}}}

\newcommand{\nAfoursixthreeb}[0]{
\parbox{100pt}{\makebox(100,40){\hspace{-15pt}
\begin{picture}(0,0)
\SetScale{1}
 \SetWidth{\thicklinewidth}
 \Line(25,15)(25,-15)
 \Line(25,-15)(40,-15) \Line(25,15)(40,15)
 \SetWidth{\thinlinewidth}
 \Line(-15,-15)(0,-15) \Line(-15,15)(25,15)
 \Line(0,-15)(15,-15) \Line(-30,-15)(25,-15)
 \Line(-15,-15)(-2,-2) \Line(2,2)(15,15)
 \Line(-15,15)(15,-15)
 \ArrowLine(-30,15)(-15,15)   \Text(-38,15)[]{$p_1$}
 \ArrowLine(-30,-15)(-15,-15) \Text(-38,-15)[]{$p_2$}
 \ArrowLine(28,15)(40,15)   \Text(49,15)[]{$p_3$}
 \ArrowLine(28,-15)(40,-15) \Text(49,-15)[]{$p_4$}
 \CArc(0,0)(10,115,155)
 \CArc(0,0)(10.1,205,245)
\end{picture}
}}}

\newcommand{\nAfoursixthreec}[0]{
\parbox{100pt}{\makebox(100,40){\hspace{-15pt}
\begin{picture}(0,0)
\SetScale{1}
 \SetWidth{\thicklinewidth}
 \Line(25,15)(25,-15)
 \Line(25,-15)(40,-15) \Line(25,15)(40,15)
 \SetWidth{\thinlinewidth}
 \Line(-15,-15)(0,-15) \Line(-15,15)(25,15)
 \Line(0,-15)(15,-15) \Line(-30,-15)(25,-15)
 \Line(-15,-15)(-2,-2) \Line(2,2)(15,15)
 \Line(-15,15)(15,-15)
 \ArrowLine(-30,15)(-15,15)   \Text(-38,15)[]{$p_1$}
 \ArrowLine(-30,-15)(-15,-15) \Text(-38,-15)[]{$p_2$}
 \ArrowLine(28,15)(40,15)   \Text(49,15)[]{$p_3$}
 \ArrowLine(28,-15)(40,-15) \Text(49,-15)[]{$p_4$}
 \CArc(0,0)(9,115,155)
 \CArc(0,0)(9.1,205,245)
 \CArc(0,0)(11,115,155)
 \CArc(0,0)(11.1,205,245)
\end{picture}
}}}

\newcommand{\nAfoursixthreemb}[0]{
\parbox{100pt}{\makebox(100,40){\hspace{-15pt}
\begin{picture}(0,0)
\SetScale{1}
 \SetWidth{\thicklinewidth}
 \Line(25,15)(25,-15)
 \Line(25,-15)(40,-15) \Line(25,15)(40,15)
 \SetWidth{\thinlinewidth}
 \Line(-15,-15)(0,-15) \Line(-15,15)(25,15)
 \Line(0,-15)(15,-15) \Line(-30,-15)(25,-15)
 \Line(-15,-15)(-2,-2) \Line(2,2)(15,15)
 \Line(-15,15)(15,-15)
 \ArrowLine(-30,15)(-15,15)   \Text(-38,15)[]{$p_1$}
 \ArrowLine(-30,-15)(-15,-15) \Text(-38,-15)[]{$p_2$}
 \ArrowLine(28,15)(40,15)   \Text(49,15)[]{$p_3$}
 \ArrowLine(28,-15)(40,-15) \Text(49,-15)[]{$p_4$}
 \GCirc(25,-10){3}{0}
 \GCirc(25,10){3}{0}
 \Text(28,3)[]{\Large \vdots}
 \Text(33,0)[]{\Huge \}}
 \Text(45,0)[]{n-1}
\end{picture}
}}}

\newcommand{\coeffaa}[1]{a_{#1}}
\newcommand{\coeffab}[1]{b_{#1}}
\newcommand{\coeffba}[1]{c_{#1}}
\newcommand{\coeffbb}[1]{d_{#1}}
\newcommand{\coeffca}[1]{e_{#1}}
\newcommand{\coeffcb}[1]{f_{#1}}
\newcommand{\coeffda}[1]{g_{#1}}
\newcommand{\coeffdb}[1]{h_{#1}}
\newcommand{\coeffea}[1]{k_{#1}}
\newcommand{\coeffeb}[1]{l_{#1}}
\newcommand{\coeffec}[1]{m_{#1}}

\makeatletter
\renewcommand\@fpheader{\hfill \parbox{3cm}{MITP/13-031\\ZU-TH~11/13\\BI-TP 2013/13}}
\renewcommand\@journal{}
\makeatother

\title{Massive planar and non--planar double box integrals\\
for light $N_f$ contributions to $gg\to t\bar{t}$}

\author[a,b]{Andreas~von~Manteuffel}
\author[c]{and Cedric~Studerus}

\affiliation[a]{
  PRISMA Cluster of Excellence \& Institute of Physics,
  Johannes Gutenberg University,
  55099 Mainz, Germany
}
\affiliation[b]{
  Institute for Theoretical Physics,
  University of Z\"urich,
  Winterthurerstrasse 190,
  8057 Z\"urich, Switzerland
}
\affiliation[c]{
  Faculty of Physics,
  University of Bielefeld,
  Postfach 100131,
  33501 Bielefeld, Germany
}

\emailAdd{manteuffel@uni-mainz.de}
\emailAdd{cedricstuderus@gmail.com}

\abstract{
We present the master integrals needed for the light fermionic
two--loop corrections to top quark pair production in the
gluon fusion channel.
Via the method of differential equations
we compute the results in terms of multiple polylogarithms
in a Laurent series about $d=4$,
where $d$ is the space--time dimension.
The most involved topology is a non--planar double box with
one internal mass.
We employ the coproduct--augmented symbol calculus and show that significant
simplifications are possible for selected results using an optimised
set of multiple polylogarithms.
}

\notoc

\begin{document}

\maketitle
\newpage


\section{Introduction}
\label{sec:intro}

\vspace*{-1mm}
Analytical calculations of next-to-next-to-leading order (NNLO) corrections
to top quark pair production at hadron colliders require, among other
ingredients, results for various two-loop Feynman integrals.
While first complete numerical NNLO predictions~\cite{Baernreuther:2012ws,Czakon:2012zr,Czakon:2012pz,Czakon:2013goa}
for the total pair production cross section have appeared recently,
in the analytical approach only a subset of the required building blocks are
available~\cite{Bonciani:2008az,Bonciani:2009nb,Bonciani:2010mn,Bernreuther:2011jt,Abelof:2011ap,Abelof:2012rv,Abelof:2012he,Korner:2008bn,Kniehl:2008fd,Anastasiou:2008vd,Bierenbaum:2011gg}
at the present time.
Here, we focus on double box master integrals
which contribute to the light fermionic corrections in the gluon channel,
i.\,e.\ all Feynman diagrams containing at least one massless
fermion in a closed loop.

The most involved integrals considered here are the three master integrals
of a particular non--planar topology with one massive propagator.
Our results for these integrals were sketched in
\cite{acat2011,vonManteuffel:2012je}.
Numerical results in the physical region of phase space have been presented in
the analysis~\cite{Borowka:2013cma} using the sector decomposition program {\tt SecDec}
\cite{Carter:2010hi,Borowka:2012yc}.
Here, we present the full analytical result and describe in more detail how we obtained it.

The outline of this paper is as follows.
In section~\ref{sec:intro}, we describe our calculational setup, which is
based on the method of differential equations~\cite{Kotikov:1990kg,Kotikov2,Kotikov3,Remiddi1,Caffo1,Caffo2,Gehrmann1,Argeri}
for a Laurent expansion in $\epsilon=(4-d)/2$.
In sections~\ref{sec:planar} and \ref{sec:nonplanar}
we present results for the planar and non--planar master integrals, respectively, given in terms
of multiple polylogarithms~\cite{Goncharov1,Goncharov2,Broadhurst1,Remiddi2,NUMHPL,Vollinga,Maitre:2005uu,Maitre:2007kp,Buehler:2011ev,Gehrmann:2000zt,Gehrmann2,harmonicsums,Ablinger:2011te,Ablinger:2013cf}.
The symbol calculus~\cite{Goncharov2,Duhr:2011zq} and its coproduct based extension~\cite{Brown:2011ik,Duhr:2012fh}
are powerful tools to exploit functional identities between multiple polylogarithms and have been succesfully applied
to both conformal theories~\cite{Goncharov:2010jf,Dixon:2011pw,Dixon:2011nj,Dixon:2012yy,Drummond:2012bg,Golden:2013xva}
and QCD~\cite{acat2011,Duhr:2012fh,Chavez:2012kn,vonManteuffel:2012je,Gehrmann:2013vga,Anastasiou:2013srw}.
For selected Laurent coefficients of our non--planar master integrals,
we employ symbol and coproduct based techniques and find remarkable simplifications.

The full result up to and including weight four, as needed for the NNLO corrections
to top quark pair production, is attached in form of a computer readable
file to the arXiv submission of this paper.
In the main text, we give the first few Laurent coefficients of the results to illustrate
their structure.


\section{Calculational method}
\label{sec:method}

\vspace*{-1mm}
Our setup for the calculation is as follows.
We identify the dimensionally regularised master integrals
required for the light fermionic two--loop corrections to $gg\to t\bar{t}$
by generating diagrams with {\tt QGRAF}~\cite{qgraf},
matching them to sectors of integral families and
reducing the loop integrals with integration-by-parts (IBP) identities
through a variant of the Laporta algorithm~\cite{Laporta1,Laporta2,Tka,Chetyrkin1}.
The last two steps are performed with {\tt Reduze\,2}~\cite{Studerus:2009ye,vonManteuffel:2012np,ginac,fermat}
(for other public reduction programs see~\cite{Anastasiou:2004vj,Smirnov:2008iw,Smirnov:2013dia}).
Ambiguities in the representation of Feynman integrals which are due to
shifts of the loop momenta, crossings of external momenta or a combination
thereof are eliminated by the program in an automated way.
For completeness, we append the definition of the integral families
we used as a file to the arXiv submission of this paper.
These integral families are also the basis for our sector naming conventions.
While many of the required master integrals are already known in the
literature~\cite{vanNeerven:1985xr,Argeri:2002wz,Bonciani:2003te,Bonciani2,Bonciani3,Aglietti:2003yc,Aglietti2,DK,Aglietti:2004tq,CGR,Bonciani:2008az,Bonciani:2009nb},
we find several sectors for which the master
integrals have not been computed in analytical form before.
These are the following.

\newpage
\noindent
For sector {\tt tt2pC:5:214}:
\begin{flalign}
\;\,
\pCtwoonefoura &\!= \int\! \frac{\mathfrak{D}^d k_1 \mathfrak{D}^d k_2}
  {D_m(k_2) D_0(k_1\!-\!k_2) D_m(k_2\!-\!p_1) D_m(k_2\!-\!p_{12}) D_0(k_1\!-\!p_3)},&&\\
\pCtwoonefourb &\!= \int\! \frac{\mathfrak{D}^d k_1 \mathfrak{D}^d k_2}
  {D_m(k_2) D_0(k_1\!-\!k_2) D_m^2(k_2\!-\!p_1) D_m(k_2\!-\!p_{12}) D_0(k_1\!-\!p_3)}.&&
\end{flalign}
\\[1ex] \noindent
For sector {\tt tt2pD:5:174}:
\begin{flalign}
\;\,
\pDonesevenfoura &\!= \int\! \frac{\mathfrak{D}^d k_1 \mathfrak{D}^d k_2}
{D_0(k_2) D_0(k_1\!-\!k_2) D_0(k_1\!-\!p_1) D_m(k_1\!+\!p_{23}) D_m(k_1\!-\!p_3)}, && \\
\pDonesevenfourb &\!= \int\! \frac{\mathfrak{D}^d k_1 \mathfrak{D}^d k_2}
{D_0(k_2) D_0(k_1\!-\!k_2) D_0(k_1\!-\!p_1) D_m^2(k_1\!+\!p_{23}) D_m(k_1\!-\!p_3)}. &&
\end{flalign}
\\[1ex] \noindent
For sector {\tt tt2pD:5:182}:
\begin{flalign}
\;\,
\pDoneeighttwoa &\!= \int\! \frac{\mathfrak{D}^d k_1 \mathfrak{D}^d k_2}
{D_0(k_2) D(k_1\!-\!k_2) D_0(k_2\!-\!p_1) D_m(k_1\!+\!p_{23}) D_m(k_1\!-\!p_3)}, &&\\
\pDoneeighttwob &\!= \int\! \frac{\mathfrak{D}^d k_1 \mathfrak{D}^d k_2}
{D_0(k_2) D(k_1\!-\!k_2) D_0(k_2\!-\!p_1) D_m(k_1\!+\!p_{23}) D_m^2(k_1\!-\!p_3)}. &&
\end{flalign}
\\[1ex] \noindent
For sector {\tt tt2pE:5:333}:
\begin{flalign}
\;\,
\pEthreethreethreea &\!= \int\! \frac{\mathfrak{D}^d k_1 \mathfrak{D}^d k_2}
 {D_0(k_1) D_m(k_1\!-\!k_2) D_0(k_1\!-\!p_1) D_0(k_2\!+\!p_{23}) D_0(k_2\!-\!p_3)},&&\\
\pEthreethreethreeb &= \int\! \frac{\mathfrak{D}^d k_1 \mathfrak{D}^d k_2}
 {D_0(k_1) D_m^2(k_1\!-\!k_2) D_0(k_1\!-\!p_1) D_0(k_2\!+\!p_{23}) D_0(k_2\!-\!p_3)}.&&
\end{flalign}
\\[1ex] \noindent
For sector {\tt tt2nA:7:463}:
\begin{flalign}
\label{nA463defa}
\;\,
\nAfoursixthreea&= \int\! \frac{\mathfrak{D}^d k_1 \mathfrak{D}^d k_2}
{D_{\text{tt2nA:7:463}}(k_1,k_2,p_1,p_2,p_3)},&&\\
\label{nA463defb}
\nAfoursixthreeb&= \int\! \frac{\mathfrak{D}^d k_1 \mathfrak{D}^d k_2\,\,\, k_1 \!\cdot\! k_2}
{D_{\text{tt2nA:7:463}}(k_1,k_2,p_1,p_2,p_3)},&&\\
\label{nA463defc}
\nAfoursixthreec&= \int\! \frac{\mathfrak{D}^d k_1 \mathfrak{D}^d k_2\,\,\, (k_1 \!\cdot\! k_2)^2}
{D_{\text{tt2nA:7:463}}(k_1,k_2,p_1,p_2,p_3)},
\end{flalign}
with the denominator
\begin{flalign}
& D_{\text{tt2nA:7:463}}(k_1,k_2,p_1,p_2,p_3) = &&\notag\\
&\hspace{6ex} D_0(k_1) D_0(k_2) D_0(k_1\!+\!p_1) D_0(k_2\!+\!p_2)
 D_0(k_1\!-\!k_2\!+\!p_1) D_0(k_1\!-\!k_2\!-\!p_2) D_m(k_1\!-\!k_2\!+\!p_{13}). &&
\raisetag{6.5ex}
\end{flalign}
The integrals involve propagator denominators with mass zero or mass $m$,
\begin{align}
\label{prop0}
 D_0(k) &= k^2 +i\delta,\qquad \\
\label{propm}
 D_m(k) &= k^2 - m^2 + i\delta\,,
\end{align}
and employ the integration measure
\begin{align}
\mathfrak{D}^d k &\equiv \frac{(2\pi)^2 m^{2\epsilon}}{C(\epsilon)} \frac{\mathrm{d}^d k}{(2 \pi)^d}, &
 C(\epsilon) &\equiv (4 \pi)^\epsilon \Gamma(1+\epsilon)\,,
\end{align}
where $d$ is the space--time dimension and $\epsilon \equiv (4-d)/2$.
The incoming momenta $p_1$ and $p_2$ fulfil $p_1^2=p_2^2=0$, while the
outgoing momenta $p_3$ and $p_4=p_1+p_2-p_3$ fulfil $p_3^2=p_4^2=m^2$.
Finally, the definitions above employ the abbreviations $p_{12}\equiv p_1 + p_2$,
$p_{13}\equiv p_1 - p_3$ and $p_{23}\equiv p_2-p_3$.
Different choices of master integrals are possible.
Our selection above leads to a (partial) decoupling of the differential equations
order by order in the Laurent expansion about $\epsilon = 0$.
This effectively allows us to solve the integrals by integrating ordinary
differential equations as we discuss in more detail below.

The four-point functions we want to compute have two massless legs and two legs with
the same non-vanishing top quark mass $m$.
Consequently, the generically 6 independent scalar products of the 3 linearly
independent external momenta reduce to 3 independent quantities in this case.
Propagators are restricted to have mass zero or $m$ and thus do not
introduce additional scales.
Therefore, all of our master integrals depend on 3 independent variables,
for which we choose $m$ and 2 dimensionless quantities out of the set $\{x,y,z\}$, where
\begin{align}
  x &= \frac{\sqrt{1-4m^2/s} - 1}{\sqrt{1-4m^2/s}+1}\,,&
  y &= -\frac{t}{m^2}\,,&
  z &= -\frac{u}{m^2}\,.
\end{align}
The Mandelstam variables are $s \equiv p_{12}^2$, $t \equiv p_{13}^2$
and $u \equiv p_{23}^2$.
The variable $x$ absorbs roots in the differential equations associated
with a massive two particle threshold, see e.g.\ \cite{Bonciani:2010ms} for
more details.
Momentum conservation implies $s+t+u=2 m^2$, which translates into
the non-linear relation
\begin{equation}
\label{mshell}
y+z =   - \frac{1+x^2}{x}
\end{equation}
for our dimensionless variables.

\begin{figure}
\centerline{\includegraphics[width=0.5\textwidth]{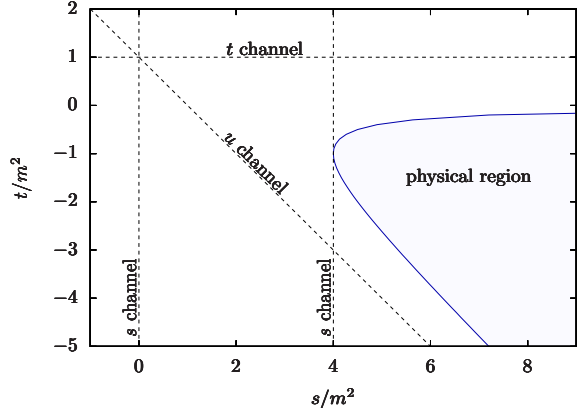}}
\caption{\label{fig:regions}
Physical region of phase space bounded by $u\,t=1$ and possible threshold singularities
at $s=0$, $s=4m^2$, $t=m^2$ and $u=m^2$.
}
\end{figure}
In the physical region of phase space for top quark pair production
the variables fulfil
\begin{align}
  m^2 &> 0,&   -1 &\leq x < 0,&   -x &\le y \le -1/x,&   -x &\le z \le -1/x,&   y z &\ge 1, & y+z &\ge 2\,.
\end{align}
Branch cut ambiguities are resolved by causality, implemented via the $i \delta$ prescription in the
Feynman propagator denominators, \eqref{prop0} and \eqref{propm}.
Depending on the topology, such a branching occurs for our master integrals due
to thresholds located at $s=0$, $s=4m^2$, $t=m^2$ or $u=m^2$,
see figure~\ref{fig:regions}.
In the physical region with $t$ and $u$ negative and $s>4m^2$ it is sufficient
to absorb these imaginary parts into an infinitesimal positive imaginary part of $s$
for the results to be well defined.
This translates to an infinitesimal positive imaginary part for $x$.
In contrast to the planar cases, solving the non--planar master integrals requires us
to take care of these prescriptions and the associated explicit imaginary parts of
transcendental functions right from the start, see section~\ref{sec:nonplanar}.
To give a well defined meaning also to all intermediate expressions we pick some reference
point in phase space, where we choose a value for $x$ with a small (but finite) positive imaginary
part and a value for $y$ with a small (positive or negative) imaginary part.
The value of $z$, including its imaginary part, is completely determined by the mass-shell
relation \eqref{mshell}, which we treat in an algebraically exact manner throughout our calculation.
Of course, our final results should not depend on arbitrary details of our intermediate
regularisation, which we also explicitly checked.

We employ the method of differential equations
to calculate the 11 unknown master integrals in analytical form.
We use {\tt Reduze\;2} to automatically calculate the differential equations,
insert the reductions and change to an alternative basis, if required.
By differentiating with respect to the overall squared scale $m^2$ we verify the correct
scaling behaviour, a feature which becomes explicit only after insertion
of the reductions.
We integrate the differential equations in the two independent dimensionless
variables and equate the solutions.
In that way we fully decouple the problem of integration from the determination
of the integration constants, which are pure numbers.
These can in principle be determined by an independent numerical evaluation
method for a couple of phase space points.
However, we prefer to give exact solutions for them.
We employ evaluations of independent Mellin-Barnes representations~\cite{Smirnov:1999gc,Tausk:1999vh}
in kinematical limits to determine
analytical expressions for the integration constants and to check the results.
For the planar topologies we used {\tt Ambre}~\cite{Gluza:2007rt} to generate
Mellin-Barnes representations, while for the non--planar topology we prepared
this representation manually, see appendix~\ref{sec:mb}.
For expansions in kinematical limits we used {\tt MB.m}~\cite{Czakon:2005rk}.
In order to determine the integration constants, we also exploit regularity
and symmetry conditions, which serves as a more convenient alternative in some cases
and as a redundant cross--check in others.
We check our results by comparing them to numerical Mellin-Barnes evaluations and
find good agreement for a choice of typically four to seven significant digits.
For the non--planar master integral~\eqref{nA463defa} we also compare
our results at different points in phase space with the numerical results
of~\cite{Borowka:2013cma} and find agreement.

Our results are expressed in terms of multiple polylogarithms.
This class of iterated integrals is defined recursively,
\begin{align}
G(w_1,\dots,w_n;x) &= \int_0^x \frac{\mathrm{d}t}{t - w_1} G(w_2,\dots,w_n;t) \qquad \text{if at least one }w_i \neq 0,\\
G(\underbrace{0,\dots,0}_{n\text{~times}};x) &= \frac{1}{n!}\ln^n(x)\,,\\
G(;x) &= 1.
\end{align}
Here, the weights $w_i\in \mathbbm{C}$, $i=1,2,\ldots,n$ and the argument $x \in \mathbbm{C}$
are considered as functions of the indeterminates.
We employ symbol and coproduct techniques
based on algorithms given in~\cite{Duhr:2011zq,Duhr:2012fh},
as well as more traditional methods for their automated treatment,
in particular for argument changes and projections onto alternative basis functions.
To implement these ideas, we have written an in-house Mathematica~\cite{mathematica} package,
which utilises the numerical evaluation implementation~\cite{Vollinga} in {\tt GiNaC}\cite{ginac}.
We emphasize that we exploit functional identities for truly complex variables
and make sure all pole prescriptions and corresponding imaginary parts are consistently
taken into account at all stages.

For univariate polylogarithms, we generalise from linear to polynomial denominators by
defining generalised weights $[f(o)]$ with
\begin{equation}
G([f(o)], w_2, \ldots, w_n; x) = \int_0^x \mathrm{d}t \frac{ f'(t)}{f(t)} G(w_2,\ldots, w_n; t)
\end{equation}
where $f(o)$ is an irreducible rational polynomial and $o$ is a dummy variable.
Without loss of generality we normalise the leading coefficient of $f$ to one.
It is curious to note that all of our integration measures with non-linear irreducible
denominators are indeed of this $\mathrm{d}\ln f(t)$ form.
A generalised weight $[f(o)]$ with the complex factorisation
\begin{equation}
 f(o) = (o-r_1)\cdots(o-r_n),
\end{equation}
where $r_i\in\mathbbm{C}$, $i=1,\ldots,n$, can be expanded in terms of standard weights according to
\begin{equation}
G(\ldots,[f(o)],\ldots; x) = G(\ldots,r_1,\ldots; x) + \ldots + G(\ldots,r_n,\ldots; x)\,.
\end{equation}
Working directly with the left hand side of this equation has the advantage that
these functions give rise to a rational symbol and do not introduce spurious imaginary parts.
While the irreducible denominators needed here are cyclotomic polynomials and
the associated cyclotomic polylogarithms defined in \cite{Ablinger:2011te} could be used
to express them, we prefer to work with the above definitions in order to emphasize
the $\mathrm{d}\ln f(t)$ structure (cf.\  \cite{Bogner:2012dn} for linear but multivariate $f$).
More details for these generalised weights will be given in another work~\cite{ManteuffelSchabinger},
where also non-cyclotomic polynomials $f(o)$ are considered.


\section{Results for planar master integrals}
\label{sec:planar}

For sector {\tt tt2pC:5:214} we choose variables $y$ and $x$ for
the differential equations and find
\begin{flalign}
\,\pCtwoonefoura &=
 -\frac{x}{m^2 (1-x)^2}
 \sum _{i=-1}^1 \coeffaa{i} \epsilon ^i + \mathcal{O}(\epsilon^2) \,, &&
\end{flalign}
\begin{subequations}
\begin{flalign}
\coeffaa{-1} &=
 -\frac{1}{32} G(0;x)^2 \hfill
&& \\
\coeffaa{0} &=
 \frac{1}{16} G(-1/x,0,-1;y)
 +\frac{1}{16} G(-x,0,-1;y)
 -\frac{1}{8} G(-1,0,-1;y)
\notag\\ &\quad
 +\frac{1}{16} G(-1/x,-1;y)G(0;x) 
 -\frac{1}{16} G(-x,-1;y)G(0;x) 
 -\frac{1}{16} G(-1/x;y) G(1,0;x)
\notag\\ &\quad
 +\frac{1}{32} G(-1/x;y) G(0;x)^2
 +\frac{1}{16} G(-x;y) G(1,0;x)
 +\frac{3}{32} G(0,1,0;x)
 +\frac{1}{4} G(0,-1,0;x)
\notag\\ &\quad
 -\frac{1}{32} G(0;x) G(1,0;x)
 -\frac{1}{48} G(0;x)^3
 +\frac{1}{48} \pi^2 G(-1/x;y)
 -\frac{1}{48} \pi^2 G(-1;y)
\notag\\ &\quad
 +\frac{1}{96} \pi^2 G(0;x)
 +\frac{3}{16} \zeta (3)
 -\frac{1}{16} G(0;x)^2\,,
\end{flalign}
\end{subequations}
\begin{flalign}
\,\pCtwoonefourb &=
 -\frac{x}{m^4 (1-x)^2}
 \sum _{i=-1}^2 \coeffab{i} \epsilon ^i + \mathcal{O}(\epsilon^3)\,, &&
\end{flalign}
\begin{subequations}
\begin{flalign}
\coeffab{-1} &= 
  \frac{1}{16} G(0;x)
  - \frac{1}{8}
  - \frac{1}{8(1-x)} G(0;x)
&& \\
\coeffab{0} &=
  - \frac{1}{16} G(-1;y) G(0;x)
  + \frac{1}{16} G(1,0;x)
  - \frac{1}{4} G(-1,0;x)
  + \frac{3}{32} G(0;x)^2
  - \frac{1}{32} \pi^2
\notag\\ &\quad
  + \frac{1}{8} G(-1;y)
  + \frac{1}{8} G(0;x)
  -\frac{3}{8}
 + \frac{1}{1-x} \Big(
    \frac{1}{8} G(-1;y) G(0;x)
  - \frac{1}{8} G(1,0;x)
\notag\\ &\quad
  + \frac{1}{2} G(-1,0;x)
  - \frac{3}{32} G(0;x)^2
  + \frac{1}{16} \pi^2
  - \frac{1}{4} G(0;x)
 \big)\,.
\end{flalign}
\end{subequations}
Since the coefficients $\coeffaa{1}$, $\coeffab{1}$ and $\coeffab{2}$
are rather lengthy we provide them only via a file on arXiv.
The solution contains multiple polylogarithms with either argument
$y$ and weights drawn from the set $\{-1, 0, -x, -1/x\}$
or with argument $x$ and weights drawn from $\{-1, 0, 1\}$.

For sector {\tt tt2pD:5:174} we choose variables $y$ and $z$ and find
\begin{flalign}
\,\pDonesevenfoura &=
 \frac{1}{m^2 (y+1)}
 \sum _{i=-1}^1 \coeffba{i} \epsilon^i + \mathcal{O}(\epsilon^2) \,, &&
\end{flalign}
\begin{subequations}
\begin{flalign}
\coeffba{-1} &=
    \frac{1}{16} G(0,-1;y)
  + \frac{1}{96} \pi^2
&& \\
\coeffba{0} &=
    \frac{1}{16} G(1/z,-1,-1;y)
  - \frac{1}{16} G(1/z,0,-1;y)
  - \frac{1}{16} G(-z-2,-1,-1;y)
\notag\\ &\quad
  - \frac{1}{16} G(1/z,-1;y)G(-1;z) 
  + \frac{1}{16} G(-z-2,-1;y)G(-1;z)
  - \frac{1}{16} G(0,-1;z) G(1/z;y)
\notag\\ &\quad
  - \frac{1}{32} G(-1;z)^2 G(-z-2;y)
  + \frac{1}{32} G(-1;z)^2 G(1/z;y)
  + \frac{3}{16} G(-1,0,-1;y)
\notag\\ &\quad
  + \frac{1}{8} G(0,0,-1;y)
  - \frac{1}{16} G(-2,-1,-1;z)
  + \frac{1}{16} G(-1,0,-1;z)
  - \frac{1}{8}  G(0,-1;y)G(-1;y)
\notag\\ &\quad
  - \frac{1}{32} \pi^2 G(-z-2;y)
  + \frac{1}{96} \pi^2 G(1/z;y)
  + \frac{1}{96} \pi^2 G(-1;y)
  - \frac{1}{32} \pi^2 G(-2;z)
\notag\\ &\quad
  + \frac{1}{96} \pi^2 G(-1;z)
  - \frac{1}{32} \pi^2 \ln 2
  + \frac{29}{64} \zeta (3)
  + \frac{1}{8} G(0,-1;y)
  + \frac{\pi^2}{48}\,,
\end{flalign}
\end{subequations}
\begin{flalign}
\,\pDonesevenfourb &=
 \frac{1}{m^4 (y+1)}
 \sum_{i=-2}^2 \coeffbb{i} \epsilon ^i + \mathcal{O}(\epsilon^3)\,, &&
\end{flalign}
\begin{subequations}
\begin{flalign}
\coeffbb{-2} &=
  \frac{1}{48}
&& \\
\coeffbb{-1} &=
  - \frac{1}{16} G(-1;y)
  - \frac{1}{48} G(-1;z)
  + \frac{5}{48}
\\
\coeffbb{0} &=
  - \frac{3}{16} G(0,-1;y)
  + \frac{1}{16} G(-1;y) G(-1;z)
  + \frac{3}{32} G(-1;y)^2
  + \frac{1}{96} G(-1;z)^2
  - \frac{7}{288} \pi^2
\notag \\ & \quad
  - \frac{1}{8} G(-1;y)
  - \frac{5}{48} G(-1;z)
  + \frac{13}{48}
 + \frac{1}{y+1} \Big(
    \frac{3}{16} G(0,-1;y)
  + \frac{1}{32} \pi^2
 \Big)\,.
\end{flalign}
\end{subequations}
We provide the coefficients $\coeffba{1}$, $\coeffbb{1}$ and $\coeffbb{2}$
via a file on arXiv.
The solution contains multiple polylogarithms with either argument
$y$ and weights drawn from the set $\{-1, 0, -2 - z, 1/z\}$
or with argument $z$ and weights drawn from $\{-2, -1, 0\}$.

For sector {\tt tt2pD:5:182} we choose variables $z$ and $y$ and find
\begin{flalign}
\,\pDoneeighttwoa &=
 \frac{1}{m^2 (y + z + 2) }
 \sum_{i=-1}^0 \coeffca{i} \epsilon^i + \mathcal{O}(\epsilon)\,,&&
\end{flalign}
\begin{subequations}
\begin{flalign}
\coeffca{-1} &=
  - \frac{1}{16} G(1/y,-1,-1;z)
  + \frac{1}{16} G(1/y,0,-1;z)
  + \frac{1}{16} G(-y-2,-1,-1;z)
&&
\notag \\ &\quad
  - \frac{1}{8} G(-1,0,-1;z)
  + \frac{1}{16} G(-2,-1,-1;y)
  - \frac{1}{8} G(-1,0,-1;y)
\notag \\ &\quad
  + \frac{1}{16} G(1/y,-1;z) G(-1;y)
  - \frac{1}{16} G(-y-2,-1;z) G(-1;y)
  + \frac{1}{16} G(0,-1;y) G(1/y;z)
\notag \\ &\quad
  + \frac{1}{32} G(-y-2;z) G(-1;y)^2
  - \frac{1}{32} G(1/y;z) G(-1;y)^2
  + \frac{1}{32} \pi^2 G(-y-2;z)
\notag \\ &\quad
  - \frac{1}{96} \pi^2 G(1/y;z)
  + \frac{1}{32} \pi^2 G(-2;y)
  - \frac{1}{48} \pi^2 G(-1;y)
  - \frac{1}{48} \pi^2 G(-1;z)
  + \frac{1}{32} \pi^2 \ln 2
\notag \\ &\quad
  - \frac{21}{64} \zeta (3)\,,
\end{flalign}
\end{subequations}
\begin{flalign}
\,\pDoneeighttwob &=
 \frac{1}{m^4 ( y + z + 2) }
 \sum_{i=-1}^1 \coeffcb{i} \epsilon^i + \mathcal{O}(\epsilon^2)\,,
&&
\end{flalign}
\begin{subequations}
\begin{flalign}
\coeffcb{-1} &=
  -\frac{1}{16} G(-1;z)
  +\frac{1}{16} G(-1;y)
 +\frac{z+1}{y+z+2} \Big(
  +\frac{1}{32} G(-1;z)^2
  -\frac{1}{16} G(-1;y) G(-1;z)
\notag \\ &\quad
  +\frac{1}{32} G(-1;y)^2
  +\frac{1}{32} \pi^2
 \Big)
&& \\
\coeffcb{0} &=
  -\frac{3}{16} G(0,-1;z)
  +\frac{3}{16} G(0,-1;y)
  +\frac{1}{8} G(-1;z)^2
  -\frac{1}{8} G(-1;y)^2
  -\frac{1}{16} G(-1;z)
\notag \\ &\quad
  +\frac{1}{16} G(-1;y)
 + \frac{z+1}{y+z+2} \Big(
   \frac{3}{16} G(1/y,-1,-1;z)
  -\frac{3}{16} G(1/y,0,-1;z)
\notag \\ &\quad
  +\frac{1}{16} G(-y-2,-1,-1;z)
  +\frac{3}{8} G(-1,0,-1;z)
  +\frac{1}{16} G(-2,-1,-1;y)
\notag \\ &\quad
  +\frac{3}{8} G(-1,0,-1;y)
  -\frac{3}{16} G(1/y,-1;z) G(-1;y)
  -\frac{1}{16} G(-y-2,-1;z) G(-1;y)
\notag \\ &\quad
  -\frac{3}{16} G(1/y;z) G(0,-1;y)
  -\frac{1}{12} G(-1;z)^3
  +\frac{1}{8} G(-1;z)^2 G(-1;y)
\notag \\ &\quad
  +\frac{3}{32} G(1/y;z) G(-1;y)^2
  +\frac{1}{32} G(-y-2;z) G(-1;y)^2
  -\frac{1}{12} G(-1;y)^3
\notag \\ &\quad
  +\frac{1}{32} \pi^2 G(1/y;z)
  +\frac{1}{32} \pi^2 G(-y-2;z)
  -\frac{1}{16} \pi^2 G(-1;z)
  +\frac{1}{32} \pi^2 G(-2;y)
\notag \\ &\quad
  -\frac{1}{16} \pi^2 G(-1;y)
  +\frac{1}{32} \pi^2 \ln 2
  +\frac{91}{64} \zeta(3)
 \Big)
 +\frac{1}{y+1} \Big(
  -\frac{3}{16} G(0,-1;y)
  -\frac{1}{32} \pi^2
 \Big)
\notag \\ &\quad
 +\frac{1}{z+1} \Big(
   \frac{3}{16} G(0,-1;z)
  +\frac{1}{32} \pi^2
 \Big)\,.
\end{flalign}
\end{subequations}
We provide the coefficients $\coeffca{0}$ and $\coeffcb{1}$
via a file on arXiv.
The solution contains multiple polylogarithms with either argument
$z$ and weights drawn from the set $\{-1, 0, -2 - y, 1/y\}$
or with argument $y$ and weights drawn from $\{-2, -1, 0\}$.

For sector {\tt tt2pE:5:333} we choose variables $y$ and $z$ and find
\begin{flalign}
\,\pEthreethreethreea &=
 \frac{1}{m^2 (y + z + 2) }
 \sum_{i=-2}^0 \coeffda{i} \epsilon^i + \mathcal{O}(\epsilon)\,, &&
\end{flalign}
\begin{subequations}
\begin{flalign}
\coeffda{-2} &=
  -\frac{1}{16} G(-1;y) G(-1;z)
  +\frac{1}{32} G(-1;y)^2
  +\frac{1}{32} G(-1;z)^2
  +\frac{\pi ^2}{32}
&& \\
\coeffda{-1} &=
   \frac{1}{4} G(-1,0,-1;y)
  +\frac{1}{8} G(-2,-1,-1;z)
  +\frac{1}{4} G(-1,0,-1;z)
  +\frac{1}{8} G(-1;y)^2 G(-1;z)
\notag \\ &\quad
  +\frac{1}{16} G(-1;z)^2 G(-z-2;y)
  +\frac{1}{16} G(-1;z)^2 G(1/z;y)
  -\frac{1}{8} G(0,-1;z) G(1/z;y)
\notag \\ &\quad
  -\frac{1}{8} G(-1;z) G(-z-2,-1;y)
  -\frac{1}{8} G(-1;z) G(1/z,-1;y)
  +\frac{1}{8} G(-z-2,-1,-1;y)
\notag \\ &\quad
  +\frac{1}{8} G(1/z,-1,-1;y)
  -\frac{1}{8} G(1/z,0,-1;y)
  -\frac{1}{12} G(-1;y)^3
  -\frac{1}{12} G(-1;z)^3
\notag \\ &\quad
  +\frac{1}{16} \pi^2 G(-z-2;y)
  +\frac{1}{48} \pi^2 G(1/z;y)
  -\frac{1}{12} \pi^2 G(-1;y)
  +\frac{1}{16} \pi^2 G(-2;z)
\notag \\ &\quad
  -\frac{1}{12} \pi^2 G(-1;z)
  +\frac{1}{16} \pi^2 \ln 2
  +\frac{35}{32} \zeta(3)\,,
\end{flalign}
\end{subequations}
\begin{flalign}
\,\pEthreethreethreeb &=
 \frac{1}{m^4 (y+1) (z+1)}
 \sum_{i=-3}^1 \coeffdb{i} \epsilon^i + \mathcal{O}(\epsilon^2) \,, &&
\end{flalign}
\begin{subequations}
\begin{flalign}
\coeffdb{-3} &=
  \frac{1}{32}
&& \\
\coeffdb{-2} &=
 -\frac{1}{16} G(-1;y)
 -\frac{1}{16} G(-1;z)
\\
\coeffdb{-1} &=
   \frac{1}{4} G(-1;y) G(-1;z)
  -\frac{1}{12}\pi^2
\\
\coeffdb{0} &=
  -\frac{3}{8} G(1/z,-1,-1;y)
  +\frac{3}{8} G(1/z,0,-1;y)
  -\frac{5}{8} G(-z-2,-1,-1;y)
\notag \\ &\quad
  -\frac{3}{8} G(-1,0,-1;y)
  -\frac{5}{8} G(-2,-1,-1;z)
  -\frac{3}{8} G(-1,0,-1;z)
\notag \\ &\quad
  +\frac{3}{8} G(1/z,-1;y) G(-1;z)
  +\frac{5}{8} G(-z-2,-1;y) G(-1;z)
  +\frac{3}{8} G(1/z;y) G(0,-1;z)
\notag \\ &\quad
  -\frac{1}{2} G(-1;y)^2 G(-1;z)
  -\frac{3}{16} G(1/z;y) G(-1;z)^2
  -\frac{5}{16} G(-z-2;y) G(-1;z)^2
\notag \\ &\quad
  -\frac{1}{16} \pi^2 G(1/z;y)
  -\frac{5}{16} \pi^2 G(-z-2;y)
  +\frac{17}{48} \pi^2 G(-1;y)
  -\frac{5}{16} \pi^2 G(-2;z)
\notag \\ &\quad
  +\frac{17}{48} \pi^2 G(-1;z)
  -\frac{5}{16} \pi^2 \ln 2
  +\frac{1}{6} G(-1;y)^3
  +\frac{1}{6} G(-1;z)^3
  -\frac{91}{32} \zeta(3)\,.
\end{flalign}
\end{subequations}
We provide the coefficients $\coeffda{0}$ and $\coeffdb{1}$
via a file on arXiv.
The solution contains multiple polylogarithms with either argument
$y$ and weights drawn from the set $\{-1, 0, -2 - z, 1/z\}$
or with argument $z$ and weights drawn from $\{-2, -1, 0\}$.


\section{Non-planar master integrals}
\label{sec:nonplanar}

The non--planar sector {\tt tt2nA:7:463} is more involved than the previous
planar cases and contains thresholds in all three channels, $s$, $t$ and
$u$.
For the integration of the differential equations, we choose the master integrals
(\ref{nA463defa}-\ref{nA463defc}).
In order to eliminate roots in $s$ from the differential equations we choose the
variable $x$ and supplement it with $y$.
Since several subsectors occur both, in their uncrossed and their
crossed version with $y\leftrightarrow z$, integrating the differential equations
with $(y,x)$ requires non-trivial argument change identities for multiple
polylogarithms with explicit imaginary parts.
We consider both kinematical invariants and master integrals to be
complex valued and keep algebraic relations between the invariants exact,
as discussed in section~\ref{sec:method}.

For the integration of the differential equations, we choose the master integrals
(\ref{nA463defa}-\ref{nA463defc}) and the variables $y$ and $x$.
As described before, we fix integration constants and check our results using
regularity constraints, symmetry conditions and a Mellin-Barnes representation.
For the Mellin-Barnes representation we choose another basis, where
the integrands contain the massive propagator to the power 1, 2 and 3, respectively,
see appendix~\ref{sec:mb}.
This Mellin-Barnes representation is described in appendix~\ref{sec:mb}.
For the solutions in the basis (\ref{nA463defa}-\ref{nA463defc}) we find
\begin{flalign}
\,\nAfoursixthreea &= \frac{x^2}{m^6 (1 - x)^2 (y+1) (1 - x + x^2 + x y)}
 \sum_{i=-4}^{0} \coeffea{i} \epsilon^i
 + \mathcal{O}(\epsilon)\,, &&
\end{flalign}
\begin{flalign}
\coeffea{-4} &=
 \frac{1}{32}
&& \\
\coeffea{-3} &=
    \frac{1}{32} G(-(1 - x + x^2)/x; y)
  - \frac{1}{32} G(-1; y)
  + \frac{1}{32} G([1 - o + o^2]; x))
  - \frac{1}{8} G(1; x)
\notag\\ &\quad
  + \frac{1}{32} G(0; x)
  + \frac{1}{32} i \pi
  + \frac{7}{96}
 + \frac{x (y+1)}{(1 - x)^2}\Big(
    \frac{1}{16} G(-(1 - x + x^2)/x; y)
  - \frac{1}{16} G(-1; y)
\notag\\ &\quad
  + \frac{1}{16} G([1 - o + o^2]; x)
  - \frac{1}{16} G(0; x)
  + \frac{1}{16} i \pi
  \Big)
\\
\coeffea{-2} &=
  - \frac{1}{32} G(-(1 - x + x^2)/x; y)^2
  - \frac{1}{16} G(-(1 - x + x^2)/x; y) G(-1; y) 
\notag\\ &\quad
  - \frac{1}{16} G(-(1 - x + x^2)/x; y) G([1 - o + o^2]; x)
  + \frac{1}{16} G(-(1 - x + x^2)/x; y) G(0; x)
\notag\\ &\quad
  - \frac{1}{32} G(-1; y)^2
  - \frac{1}{16} G(-1; y) G([1 - o + o^2]; x)
  + \frac{1}{4} G(-1; y) G(1; x)
\notag\\ &\quad
  - \frac{1}{16} G(-1; y) G(0; x)
  + \frac{1}{16} G([1 - o + o^2]; x) G(0; x)
  - \frac{1}{32} G([1 - o + o^2]; x)^2
\notag\\ &\quad
  + \frac{1}{8} G(1; x)^2
  - \frac{1}{8} G(1; x) G(0; x)
  - \frac{1}{16} i \pi G(-(1 - x + x^2)/x; y)
  - \frac{1}{16} i \pi G(-1; y)
\notag\\ &\quad
  - \frac{1}{16} i \pi G([1 - o + o^2]; x)
  + \frac{1}{16} i \pi G(0; x)
  - \frac{7}{192} \pi^2
  + \frac{1}{8} G(-(1 - x + x^2)/x; y)
\notag\\ &\quad
  - \frac{1}{6} G(-1; y)
  + \frac{1}{8} G([1 - o + o^2]; x)
  - \frac{1}{4} G(1; x)
  + \frac{1}{8} i \pi
  - \frac{7}{24}
 + \frac{ x (y+1)}{(1 - x)^2} \Big(
   \Big(
     \frac{1}{4} G(1; x)
\notag\\ &\quad
   - \frac{1}{8} G(0; x)
   - \frac{7}{24}
   \Big)\Big(
     - G(-(1 - x + x^2)/x; y)
     + G(-1; y)
     - G([1 - o + o^2]; x)
\notag\\ &\quad
     + G(0; x)
     - i \pi
   \Big)
 \Big)\,,
\end{flalign}
\begin{flalign}
\,\nAfoursixthreeb &= \frac{x^2}{m^4 (1 - x)^2 (1 - x + x^2 + x y)}
 \sum_{i=-4}^{0} \coeffeb{i} \epsilon^i
 + \mathcal{O}(\epsilon)\,, &&
\end{flalign}
\begin{flalign}
\coeffeb{-4} &=
  \frac{7}{384}
&& \\
\coeffeb{-3} &=
  - \frac{5}{192} G(-(1 - x + x^2)/x; y)
  + \frac{1}{64} G(-1; y)
  - \frac{5}{192} G([1 - o + o^2]; x)
  - \frac{1}{16} G(1; x)
\notag\\ &\quad
  + \frac{11}{192} G(0; x)
  - \frac{5}{192} i \pi
\\
\coeffeb{-2} &=
  + \frac{1}{192} G(-(1 - x + x^2)/x; y)^2
  - \frac{1}{32}  G(-(1 - x + x^2)/x; y) G(-1; y)
\notag\\ &\quad
  + \frac{1}{96} G(-(1 - x + x^2)/x; y) G([1 - o + o^2]; x)
  + \frac{1}{8}  G(-(1 - x + x^2)/x; y) G(1; x)
\notag\\ &\quad
  - \frac{7}{96} G(-(1 - x + x^2)/x; y) G(0; x)
  - \frac{1}{64} G(-1; y)^2
  - \frac{1}{32} G(-1; y) G([1 - o + o^2]; x)
\notag\\ &\quad
  + \frac{1}{32} G(-1; y) G(0; x)
  + \frac{1}{192} G([1 - o + o^2]; x)^2
  + \frac{1}{8} G([1 - o + o^2]; x) G(1; x)
\notag\\ &\quad
  - \frac{7}{96} G([1 - o + o^2]; x) G(0; x)
  + \frac{1}{16} G(1; x)^2
  - \frac{3}{16} G(0; x) G(1; x)
  + \frac{1}{12} G(0; x)^2
\notag\\ &\quad
  + \frac{1}{96} i \pi G(-(1 - x + x^2)/x; y)
  - \frac{1}{32} i \pi G(-1; y)
  + \frac{1}{96} i \pi G([1 - o + o^2]; x)
\notag\\ &\quad
  + \frac{1}{8}  i \pi G(1; x)
  - \frac{7}{96} i \pi G(0; x)
  - \frac{35}{1152} \pi^2
\end{flalign}
\begin{flalign}
\,\nAfoursixthreec &= -\frac{x}{m^2 (1 - x + x^2 + x y)}
 \sum_{i=-4}^{0} \coeffec{i} \epsilon^i
 + \mathcal{O}(\epsilon)\,, &&
\end{flalign}
\begin{subequations}
\begin{flalign}
\coeffec{-4} &=
   \frac{1}{256}
&& \\
\coeffec{-3} &=
  -\frac{1}{64} G(1;x)
  +\frac{1}{128} G(0;x)
  -\frac{1}{192}
\\
\coeffec{-2} &=
   \frac{1}{32} G(1;x)^2
  -\frac{1}{32} G(1;x) G(0;x)
  +\frac{1}{128} G(0;x)^2
  -\frac{1}{192}\pi^2
\notag \\ &\quad
  +\frac{1}{48} G(-(1-x+x^2)/x;y)
  +\frac{1}{48} G([1-o+o^2];x)
  -\frac{1}{48} G(0;x)
  +\frac{1}{48} i \pi
  -\frac{1}{96}
\notag \\ &\quad
 -\frac{x (y+1)}{128(1-x)^2} \Big(
   G(-(1-x+x^2)/x;y)
  -G(-1;y)
  +G([1-o+o^2];x)
  -G(0;x)
\notag \\ &\quad
  +2 i \pi
 \Big) \Big(
  -G(-(1-x+x^2)/x;y)
  +G(-1;y)
  -G([1-o+o^2];x)
  +G(0;x)
 \Big)
\end{flalign}
\end{subequations}
We provide the coefficients $\coeffea{-1}$, $\coeffea{0}$,
$\coeffeb{-1}$, $\coeffeb{0}$,
$\coeffec{-1}$ and $\coeffec{0}$
via a file on arXiv.
We remark that our solutions for the finite terms contain exact numbers
for all integration constants except for one constant,
for which we supply a numerical approximation only.
It turns out that the light $N_f$ contributions to $gg\to t\bar{t}$ 
at NNLO are actually independent of this constant.
The full solution contains multiple polylogarithms with either argument
$y$ and weights drawn from the set
$\{-1, 0, -x, -1/x, -(1 + x^2)/x, -(1 - x + x^2)/x\}$
or with argument $x$ and weights drawn from
$\{-1, 0, 1, [1 + o^2], [1 - o + o^2]\}$.

Significant simplifications are possible for all poles in $\epsilon$ of the
first two master integrals, including the $1/\epsilon$ terms not displayed
above because of their length.
The $1/\epsilon$ pole of the third master integral and the finite parts are
considerably more involved and therefore omitted in the following.
Guided by the symbol we construct a new set of multiple polylogarithms
which we can express most naturally with the variables $y_1=y+1$ and
$z_1=z+1$.
Using the coproduct extended symbol calculus we obtain for the first
master integral the simplified expressions
\begin{flalign}
\coeffea{-4} &=
 \frac{1}{32}
&& \\
\coeffea{-3} &=
 -\frac{1}{16} \ln(y_1+z_1)
 +\frac{1}{16} i \pi 
 +\frac{7}{96}
 +\frac{y_1-z_1}{32(y_1+z_1)} \ln(y_1/z_1)
\notag \\
\coeffea{-2} &=
 -\frac{1}{32} \ln^2(y_1 z_1)
 +\frac{1}{32} \ln^2(y_1+z_1)
 +\frac{1}{16} \ln(y_1+z_1) \ln (y_1 z_1)
 -\frac{1}{16} i \pi  \ln(y_1 z_1)
\notag \\ &\quad
 -\frac{1}{16} i \pi  \ln (y_1+z_1)
 -\frac{19}{192} \pi^2
 -\frac{1}{48} \ln(y_1 z_1)
 -\frac{1}{8} \ln(y_1+z_1)
 +\frac{1}{8} i \pi
 -\frac{7}{24}
\notag \\ &\quad
 +\frac{y_1-z_1}{y_1+z_1} \ln(y_1/z_1) \Big(
   -\frac{1}{16} \ln(y_1+z_1) 
   +\frac{1}{16} i \pi
   +\frac{7}{48}
 \Big)
\\
\coeffea{-1} &=
 -\frac{1}{16} G\Big(1,0,0;\frac{y_1 z_1}{y_1+z_1}\Big)
 -\frac{1}{16} i \pi  G\Big(1,0;\frac{y_1 z_1}{y_1+z_1}\Big)
 +\frac{1}{48} \ln^3(y_1 z_1)
\notag \\ &\quad
 -\frac{1}{16} \ln^2(y_1+z_1) \ln(y_1 z_1)
 +\frac{1}{8} i \pi  \ln(y_1+z_1) \ln(y_1 z_1)
 +\frac{3}{32} \pi^2 \ln(y_1 z_1)
\notag \\ &\quad
 +\frac{5}{48} \pi^2 \ln(y_1+z_1)
 -\frac{29}{32} \zeta(3)
 -\frac{5}{48} i \pi^3
 -\frac{5}{48} \ln^2(y_1 z_1)
 +\frac{7}{48} \ln^2(y_1/z_1)
\notag \\ &\quad
 +\frac{1}{4} \ln(y_1+z_1) \ln(y_1 z_1)
 -\frac{1}{4} i \pi  \ln(y_1 z_1)
 +\frac{1}{12} \ln(y_1 z_1)
 +\frac{1}{2} \ln(y_1+z_1)
\notag \\ &\quad
 -\frac{1}{144} \pi^2 
 -\frac{1}{2} i \pi
 +\frac{7}{6}
 +\frac{y_1-z_1}{y_1+z_1} \ln(y_1/z_1) \Big(
  -\frac{1}{48} \ln^2(y_1/z_1)
  +\frac{1}{16} \ln^2(y_1+z_1)
\notag \\ &\quad
  -\frac{1}{8} i \pi \ln(y_1+z_1)
  -\frac{17}{96} \pi^2
  -\frac{1}{24} \ln(y_1 z_1)
  -\frac{1}{4} \ln(y_1+z_1)
  +\frac{1}{4} i \pi
  -\frac{7}{12}
 \Big)
\end{flalign}
Here we choose a representation which makes the forward-backward symmetry
$y_1 \leftrightarrow z_1$ of the corner integral explicit.
For the second master integral we find
\begin{flalign}
\coeffeb{-4} &= 
 \frac{7}{384}
&& \\
\coeffeb{-3} &= 
 -\frac{1}{32} \ln(y_1+z_1)
 +\frac{1}{64} \ln y_1
 -\frac{5}{192}  \ln  z_1
 +\frac{1}{32} i \pi
\\
\coeffeb{-2} &= 
 \frac{1}{64} \ln^2(y_1+z_1)
 -\frac{1}{64} \ln^2 y_1
 +\frac{1}{192} \ln^2 z_1
 -\frac{1}{32} \ln  y_1 \ln z_1
\notag\\ &\quad
 +\frac{1}{16} \ln  z_1 \ln(y_1+z_1)
 -\frac{1}{32} i \pi  \ln(y_1+z_1)
 -\frac{1}{16} i \pi  \ln z_1
 -\frac{47}{1152} \pi^2
\\
\coeffeb{-1} &= 
 -\frac{1}{32} G\Big(1,0,0;\frac{y_1 z_1}{y_1+z_1}\Big)
 -\frac{1}{32} i \pi  G\Big(1,0;\frac{y_1 z_1}{y_1+z_1}\Big)
 +\frac{1}{16} \ln ^2 y_1 \ln  z_1
\notag\\ &\quad
 -\frac{1}{16} \ln  z_1 \ln^2(y_1+z_1)
 +\frac{1}{8} i \pi  \ln  z_1 \ln(y_1+z_1)
 +\frac{5}{96} \pi^2 \ln(y_1+z_1)
\notag\\ &\quad
 -\frac{1}{24} \pi^2 \ln  y_1
 -\frac{1}{144} \ln^3 z_1
 +\frac{29}{288} \pi^2 \ln  z_1
 -\frac{55 \zeta(3)}{192}
 -\frac{5}{96} i \pi^3
\end{flalign}
The original expressions for these poles in terms of $G$ functions with
argument $y$ or $x$ contained 65 multiple polylogarithms
(22 two-dimensional and weight $> 1$) when
all products are expanded with the shuffle relations.
Systematically exploiting relations between them by a coproduct based
reduction procedure reduces this number to 28 multiple polylogarithms
(12 two-dimensional and weight $>1$).
This is reduced by an optimised choice of basis functions to just
$\Li_3(y_1 z_1/(y_1+z_1))$, $\Li_2(y_1 z_1/(y_1+z_1))$,
$\ln(y_1+z_1)$, $\log y_1$ and
$\ln z_1$.
Note that in the above expressions we used a more compact $G$ function
based notation, which can easily be converted to classical polylogarithms
via $\Li_3(x)=-G(0,0,1;x)$, $\Li_2(x)=-G(0,1;x)$ and shuffle relations.
Finally, we remark that the original expressions contained roots in $s$
through $x$, both in the rational prefactors and in the multiple polylogarithms,
which could all be eliminated in the above expressions.

\section{Conclusions}
\label{sec:conc}

In this work, we presented analytical solutions for double box master integrals,
which have not been available before.
For the first time, we gave explicit solutions for non--planar
double box integrals with a massive propagator in terms of multiple
polylogarithms.
Our results complete the set of master integrals required
for the analytical calculation~\cite{lightnf} of the
light $N_f$ corrections to $gg\to t\bar{t}$ at the two-loop level.

By carrying out a coproduct--augmented symbol analysis of the poles of two
non--planar master integrals we demonstrated that remarkable simplifications
are possible using an optimised set of multiple polylogarithms.
It has been shown in~\cite{Henn:2013pwa,Henn:2013tua} for the case of
massless, planar two--loop and three--loop four--point topologies that it is possible
to choose a basis in which the differential equations for the master integrals
take a special and particularly simple form.
In this basis, the master integrals have uniform transcendentality and
no algebraic prefactors.
Applying this method to the integrals discussed in this paper and choosing
an appropriate set of multiple polylogarithms should allow to rewrite the full
set of solutions in a very compact form~\cite{nA463uniform}.


\acknowledgments{
We thank Claude Duhr, Andrea Ferroglia and Erich Weihs for helpful discussions
on the coproduct--augmented symbol calculus, Robert Schabinger for constructive exchange on the
generalised weights and comments on the draft,
Roberto Bonciani and Lorenzo Tancredi
for interesting discussions on the method of differential equations,
Thomas Gehrmann for useful discussions and comments on the draft,
and Gudrun Heinrich and Sophia Borowka for help with the comparison of numerical results.
The work of A.\ v.\ M.\ was supported in part by the Schweizer Nationalfonds
(Grant 200020\_124773/1), by the Research Center {\em
Elementary Forces and Mathematical Foundations (EMG)} of the
Johannes Gutenberg University of Mainz and by the
German Research Foundation (DFG).
All figures were drawn with {\tt Axodraw} \cite{axodraw}.
}


\appendix


\section{A Mellin-Barnes representation for sector {\tt tt2nA:7:463}}
\label{sec:mb}

In this appendix, we give a Mellin-Barnes representation for integrals
of the non--planar sector {\tt tt2nA:7:463} where the integrand contains
the massive propagator taken to the power $n$.
Similar to the calculation~\cite{Tausk:1999vh} in the massless case,
we start from a Feynman parameter representation, integrate the Feynman
parameters at the expense of introducing Mellin-Barnes contour integrals
and obtain
\begin{flalign}
\,
&\nAfoursixthreemb = \int\! \frac{\mathfrak{D}^d k_1 \mathfrak{D}^d k_2}
{D_{\text{tt2nA:7:463}}(k_1,k_2,p_1,p_2,p_3)\;D^{n-1}_m(k_1\!-\!k_2\!+\!p_{13})}
&&\\
&\;=
\frac{m^{8-2d} \mu^{-4-2n} (-1)^{n - 1} \Gamma(-2 + d/2)^2}
{16 \Gamma^2(3-d/2)\Gamma(n) \Gamma(-4 + d) \Gamma(-6 - n + 3 d/2)}
\int_{\mathcal{C}_1}\! \frac{\mathrm{d} z_1}{2\pi i}
\int_{\mathcal{C}_2}\! \frac{\mathrm{d} z_2}{2\pi i}
\int_{\mathcal{C}_3}\! \frac{\mathrm{d} z_3}{2\pi i}
\int_{\mathcal{C}_4}\! \frac{\mathrm{d} z_4}{2\pi i}
\int_{\mathcal{C}_5}\! \frac{\mathrm{d} z_5}{2\pi i}
&&\notag\\ &\quad
\Big(\frac{-s}{\mu^2}\Big)^{-6 - n + d - z_1 - z_2 - z_5} \Big(\frac{-t_1}{\mu^2}\Big)^{z_1} \Big(\frac{-u_1}{\mu^2}\Big)^{z_2}
 \Big(\frac{m^2}{\mu^2}\Big)^{z_5}\;
 \frac{\Gamma(-z_1) \Gamma(-z_2) \Gamma(-z_3) \Gamma(-z_4)  \Gamma(-z_5)}
 {\Gamma^2(2 + z_1 + z_2 + z_3 + z_4)}
&&\notag\\ &\quad
\Gamma(1 + z_1 + z_3) \Gamma(1 + z_2 + z_3)
 \Gamma(1 + z_1 + z_4) \Gamma(1 + z_2 + z_4)
\Gamma(n + z_1 + z_2 + 2 z_5)
&&\notag\\ &\quad
    \Gamma(
    4 - d/2 + z_1 + z_2 + z_3 + z_4) \Gamma(-5 - n + d - z_1 - z_2 - z_3 - 
     z_5)
&&\notag\\ &\quad
\Gamma(-5 - n + d - z_1 - z_2 - z_4 - z_5)
     \Gamma(6 + n - d + z_1 + z_2 + z_3 + z_4 + z_5)
\end{flalign}
where $t_1=t-m^2$, $u_1=u-m^2$ and $\mu$ is an auxiliary normalisation scale.
The contours $\mathcal{C}_1,\ldots,\mathcal{C}_5$ of complex integration
are for imaginary parts from $-\infty$ to $+\infty$ and, for simplicity,
fixed real parts choosen to separate the towers of increasing and decreasing
poles of the different $\Gamma$ functions.
Despite the fact that this representation requires only one contour
integration more than in the massless case, its evaluation is significantly
more involved.


\end{document}